\documentclass[aps,pre,superscriptaddress,multicol,showkeys]{revtex4}
\usepackage{amsmath,amssymb,amsfonts,latexsym}
\usepackage{mathrsfs}
\usepackage{rotating}

\begin{document}

\title{ Symmetric L\'evy flights in Semi-infinite domain}
\author{Barnali Pyne  }
\email{barnalipyne22@gmail.com}
\affiliation{ Department of Physics, Ramniranjan Jhunjhunwala College, Ghatkopar (West), Mumbai 400086, India }
\affiliation{Department of Mathematics, IIEST, Shibpur, Botanic Garden, Howrah-711103, West Bengal, India\footnote{Present address}}
\author{Kiran M. Kolwankar}
\email{Kiran.Kolwankar@rjcollege.edu.in}
\affiliation{ Department of Physics, Ramniranjan Jhunjhunwala College, Ghatkopar (West), Mumbai 400086, India }

%\date{}
% \vspace{1cm}
%\maketitle  \centerline{{\bf ABSTRACT}}
 % \vspace {0.2 cm}
\begin{abstract}

We study symmetric L\'evy flights in a semi-infinite domain $[0.\infty)$ with a reflecting and absorbing boundary at 0. To this end, we use the fractional differential equation that governs the L\'evy process. Incorporating the boundary conditions in L\'evy flights has been an open and tricky question, as the long jumps can lead to the L\'evy flights leaping over the boundary. We, for the first time, incorporate reflecting and absorbing boundary conditions for L\'evy flights and solve the fractional differential equation analytically to find the probability densities. Monte Carlo simulations are also performed for both boundary conditions to verify the results. Analytical and simulation results perfectly coincide for the reflecting boundary condition and, for the absorbing boundary condition, they coincide for the large abscissa values.
   
\end{abstract}
%  \thispagestyle{empty}

%\vspace {0.3 cm}


\maketitle

\section{ Introduction}

L\'evy flights~\cite{zasla, klafter1986} offer a useful mathematical model to understand phenomena in several fields that include financial marketing, cryptography, astronomy, biology, physics and nowadays even in social media.  L\'evy flights have their origin in diffusion processes. In particular, they show anomalous diffusion~\cite{metkla, klafter96}. L\'evy flights stand for a class of non-Gaussian random processes whose stationary increments are distributed according to a L\'evy stable distribution with infinite variance, originally studied by French mathematician Paul Pierre L\'evy. They can describe all scale-invariant stochastic processes~\cite{mandelbrot}. L\'evy flights have application in a wide variety of fields such as the distribution of human travel~\cite{hufnagel2006, hongchong}, describing animal foraging patterns~\cite{ klafter2015, Catalan, metzler} and even in some aspects of earthquake dynamics~\cite{ corral, costa}. Transport based on L\'evy flights has been extensively studied numerically~\cite{HH, wiersma, geisel}. The hopping processes along a polymer~\cite{sokolov, ott} can have long jumps as some chemically remote segments can come into close contact in the embedding space due to looping. The spread of diseases~\cite{hufnagel2006, hufnagel2004} due to the high degree of connectivity of air traffic from remote geographic locations involves L\'evy travel lengths and thus causes a fast worldwide spread. 

Due to several applications, L\'evy flights have received a lot of attention for more than three decades. Normally, L\'evy flights are studied in an infinite domain. However, practical problems can involve some boundaries, and the effect of a boundary on the L\'evy flight is still intriguing. Few authors have shed some light on this field in recent years~\cite{garbaczewski2018, garbaczewski2019, nowak2017, deng, Araujo}. But still, the analytical formulation and implementation of reflecting as well as absorbing boundary conditions for L\'evy flights are still unsolved. Taking into account the effect of a boundary poses problems because there is a possibility that a walker performing the L\'evy flight can jump over the boundary. This causes great difficulty in relating the models to the experimental data, especially when analyzing the scaling of the measured moments in time. The method of images has been found to fail in the case of L\'evy flights~\cite{dybiec}. Questions also arise regarding the non-uniqueness of the density of the first passage time~\cite{SM}. Mathematically, for normal diffusion, the absorbing boundary is incorporated into the diffusion equation demanding that the probability density is zero at the boundary, and for a reflecting boundary, it is the derivative of the probability density that is zero at the boundary~\cite{chandrasekhar}. It is known that the anomalous diffusion governed by the L\'evy distribution is described by a fractional-order differential equation instead of an ordinary one. The main aim of the present paper is to incorporate the absorbing and the reflecting boundary condition for symmetric L\'evy flights in the governing fractional differential equation \cite{podlubny} and find the resulting probability density analytically. In doing so, we use the methods developed in~\cite{schneider1,schneider2,mainardi2001,mainardi2005,MPG}. We also verify the analytical results by numerical simulations of symmetric L\'evy flights in the semi-infinite domain for both boundary conditions.

The paper is organized as follows. In Section II we define and discuss the effect of reflecting boundary condition for symmetric L\'evy flights analytically and furthermore compare these boundary effects explicitly with numerical simulations for the same. In Section III, the absorbing boundary condition for symmetric L\'evy flights is explored. In Section IV we conclude after discussing a few possible consequences of our results.

\section{Reflecting boundary}

The purpose of this section is to evolve a way to incorporate a reflecting boundary for L\'evy flight motion into the equation that describes it. Stable processes in a bounded domain with reflections have been considered in~\cite{BK1,BK2}. The way in which reflection is implemented there resembles resetting in the sense that when the walker crosses the boundary, it is reintroduced in the domain with some probability distribution. As we shall see in the following, we use the probability current across the boundary to incorporate the reflection. It is known that the symmetric L\'evy process is described by a fractional differential equation given by
\begin{equation}\label{eqR1}
\frac{\partial}{\partial t}P(x,t)= ~_{-\infty}D_x^{\alpha}P(x,t) + ~_xD_{\infty}^{\alpha}P(x,t) 
\end{equation}
where the Weyl fractional derivatives \cite{oldhum} on the RHS are given by
\begin{widetext}
\begin{equation}\label{WFD1}
_{-\infty}D_x^{\alpha}P(x,t)=\frac{1}{\Gamma(2-\alpha)}\frac{\partial^2}{\partial x^2}\displaystyle \int_{-\infty}^x \frac{P(y,t)}{(x-y)^{\alpha-1}}dy \;\;\;\; 1 <\alpha< 2
\end{equation}
and
\begin{equation}\label{WFD2}
_xD_{\infty}^{\alpha}P(x,t)=\frac{1}{\Gamma(2-\alpha)}\frac{\partial^2}{\partial x^2}\displaystyle \int_x^\infty \frac{P(y,t)}{(y-x)^{\alpha-1}}dy \;\;\;\; 1 <\alpha< 2.
\end{equation}
\end{widetext}
In order to arrive at an appropriate boundary condition for a reflecting L\'evy flight, we write Eq.~({\ref{eqR1}}) as
\begin{equation}\nonumber 
 \frac{\partial}{\partial t} P(x,t)=\frac{\partial}{\partial x} S(x,t)
\end{equation}
which is nothing but a continuity equation and the probability current, $S(x,t)$, is given by
\begin{widetext}
\begin{equation}\label{RBC}
S(x,t)=\frac{1}{\Gamma(2-\alpha)}\frac{\partial}{\partial x} \left\{ \displaystyle\int_{-\infty}^x \frac{P(y,t)}{(x-y)^{\alpha-1}}dy +\displaystyle\int_x^\infty \frac{P(y,t)}{(y-x)^{\alpha-1}}dy \right\}
\end{equation}
\end{widetext}
As the reflecting boundary is characterized by zero probability current across the boundary, the correct way to incorporate the reflecting boundary mathematically would be to demand that
\begin{equation}\label{RBC1}
S(0,t)=0       \;\;\;\; \forall \;t
\end{equation}

The next task is to solve Eq.~(\ref{eqR1}) subject to this boundary condition and obtain an expression for the resultant probability density. 
We have carried it out (see Appendix A) by assuming that $P(x,t)$ is an even function and hence making use of the Fourier cosine transform. We considered the initial condition  $P(x,0)=\delta(x-a)$ where initially the particles are on $x = a $ with $a\geq0$ and arrived at the following integral:
\begin{equation}\label{eqR2}
P(x,t) = \frac{2}{\pi}\displaystyle\int_0^\infty e^{-Ak^\alpha t} \cos(ka)\cos(kx)dk 
\end{equation}
where $A=2\cos(\frac{\pi}{2}(2-\alpha))$.
In order to solve this integral, we used the Mellin transform~\cite{mainardi2001,mainardi2007,mainardi2010} and the fundamental result on the Mellin Barnes integral~\cite{braaksma, paris}. Then by solving the complex integral (Appendix B), we obtained the following expression for $P(x,t)$:
\begin{widetext}
\begin{equation}\label{eqR3}
P(x,t) = \frac{1}{\pi}  \sum_{n = 1}^{\infty} \frac{(-1)^{n+1}}{n!} t^{-\frac {n}{\alpha}} A^{-\frac{n}{\alpha}}\Gamma\left(1+\frac{n}{\alpha}\right) \sin\left(\frac{n\pi}{2}\right) \left[(x+a)^{(n-1)}+(x-a)^{(n-1)}\right]
\end{equation}
\end{widetext}
This solution can also be represented by the general Fox function \cite{schneider1, schneider2, mainardi2005} ,
\begin{equation}\label{eqR4}
P(x,1/A)=\frac{1}{\alpha(x+a) }H^{11}_{22} \left( x+a \Big|^{(1,\frac{1}{\alpha}),(1,\frac{1}{2})}_{(1,1),(1,\frac{1}{2})}\right)+\frac{1}{\alpha(x-a) }H^{11}_{22} \left( x-a \Big|^{(1,\frac{1}{\alpha}),(1,\frac{1}{2})}_{(1,1),(1,\frac{1}{2})}\right)
\end{equation}

The solution  given by (\ref{eqR3}) for $a=0$ can be written \cite{bjwest}, for the large value of x, as:
\begin{equation}\label{eqR5}
P(x,t)\sim \frac{2}{\pi} At \Gamma(1+\alpha) \sin(\frac{\pi\alpha}{2}) x^{-(1+\alpha)}.
\end{equation} 
This shows that the tail part of the curve decays as $ x^{-(1+\alpha)}$,  which is the same as the L\'evy flight without the boundary.
%\subsection{Numerical Result}

To verify this result, the L\'evy flight random walk with reflecting boundary was simulated using the GSL library function for the random numbers distributed according to the L\'evy distribution:
\begin{figure}[ht]
          \centering
          \includegraphics[width=1.0\columnwidth]{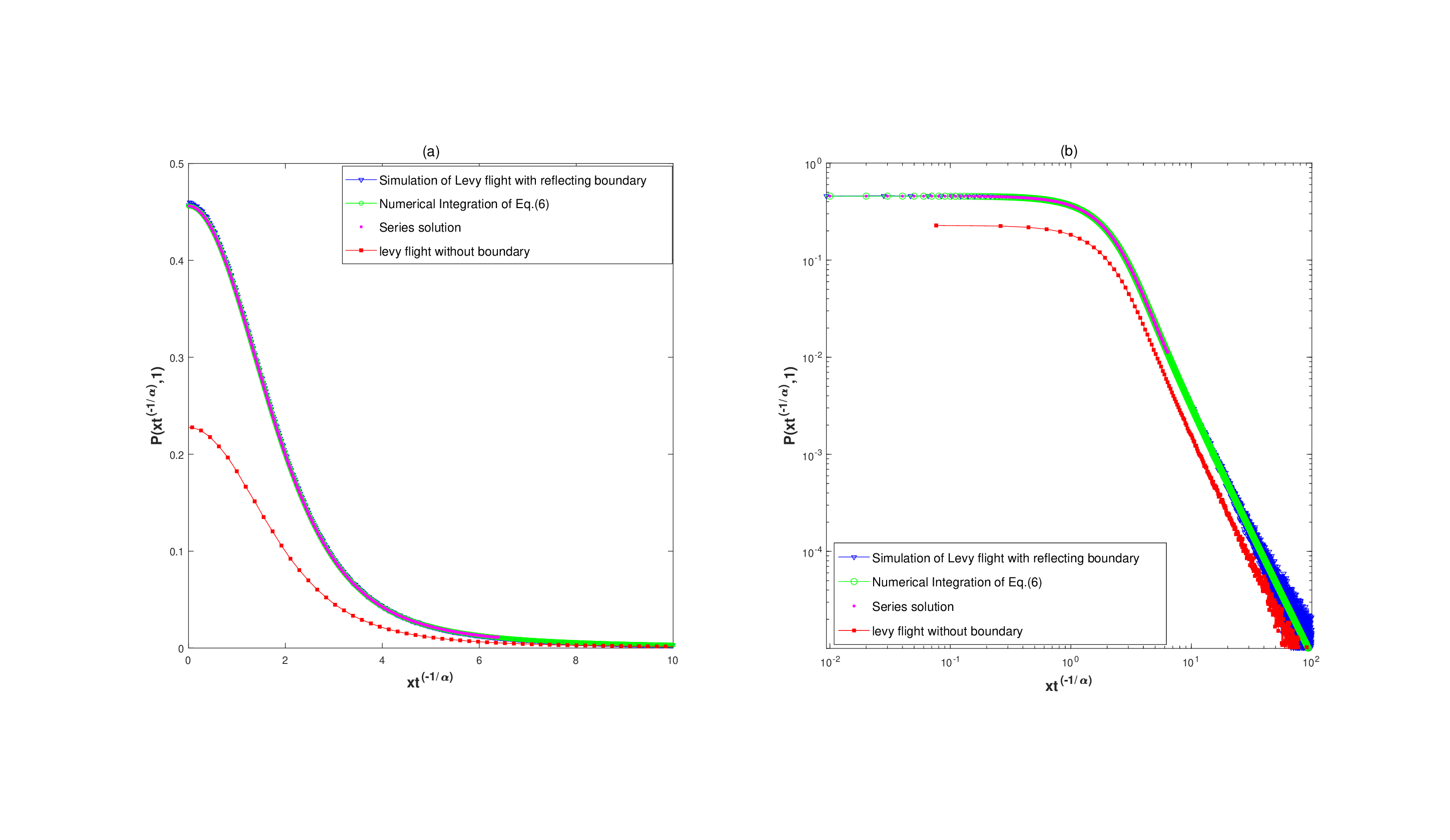}     
\caption{ Comparison of the solution given by the numerical integration of the equation (\ref{eqR2}) (green circles), series solution provided by equation (\ref{eqR3}) (pink dots) and the simulation of the probability distribution function $P(x,t)$  of L\'evy flight with reflecting boundary (blue triangles). All three plots overlap nicely. The plot for the simulation of L\'evy flight without boundary (red squares) is also shown for comparison.  ($~\alpha=1.5~$): (a) linear scale (b) log scale. It is clear that there is exact match between the first three plots and, as expected, the fourth plot is different and lower.}
      \end{figure}

\begin{figure}[ht]
  \centering
          \includegraphics[width=1.0\columnwidth]{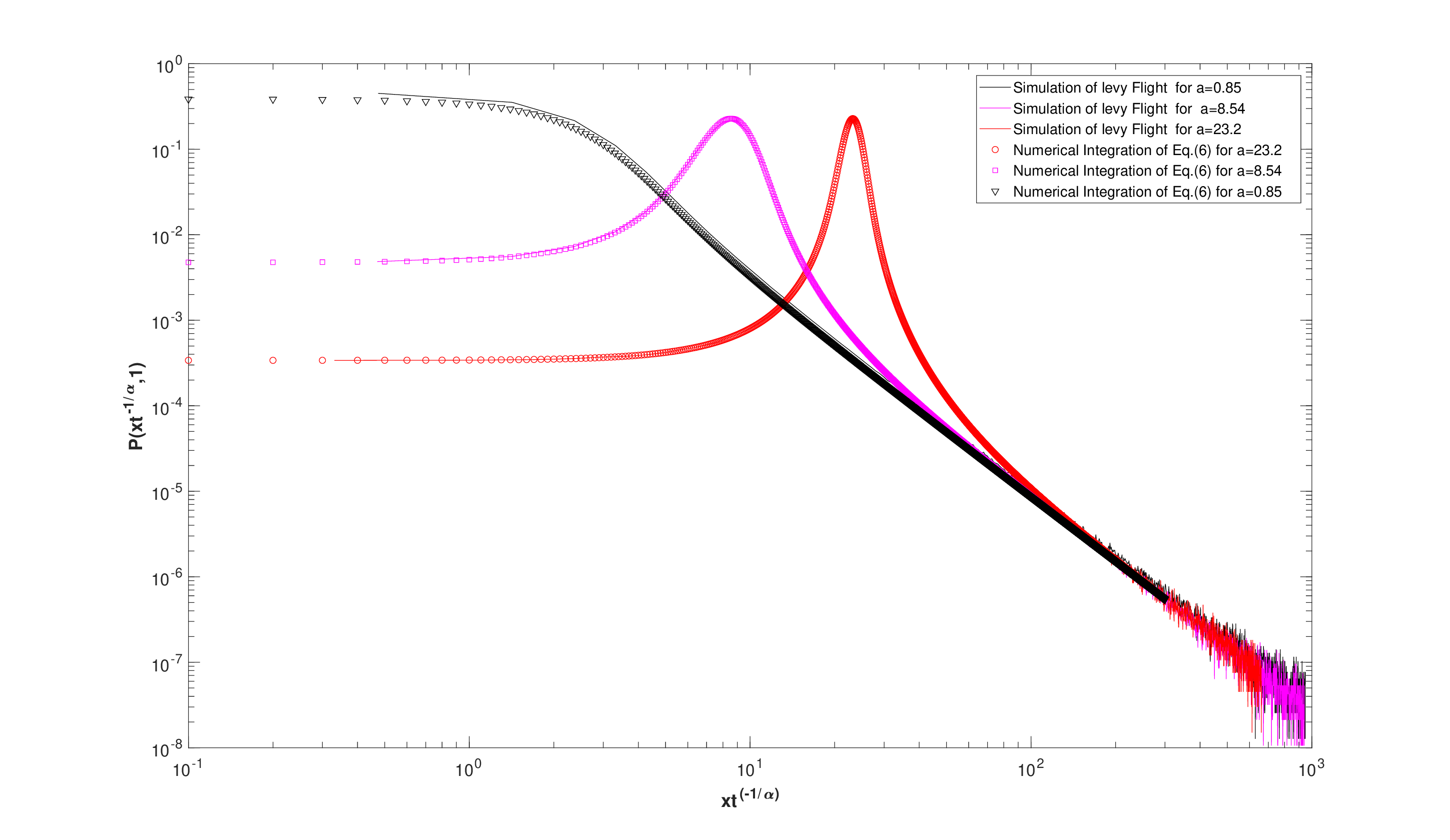}     
\caption{ Comparison of the solution given by the numerical integration of the equation (\ref{eqR2}) as well as the simulation of the probability distribution function $P(x,t)$  of L\'evy flight with reflecting boundary for the value $~\alpha=1.5~$ for different initial positions (i) $a=0.85$ (black triangles), (ii) $a=8.54$ (pink squares) and (iii) $a=23.2$ (red circles). The solid lines are corresponding simulation results. It is clear that the analytical and simulation results coincide.}
      \end{figure}

\begin{equation}\label{eqR6}
P(x) = \frac{1}{2\pi}\displaystyle\int_{-\infty}^\infty exp(-ikx-|ck|^\alpha) dk . 
\end{equation}
The value of $c$ here is $|A|^{1/\alpha}$.
It can be taken as a measure of the spread of the density similar to the second moment of a Gaussian. 
The reflecting boundary  wall  was placed at $x=0$, i. e., whenever the walker crossed the boundary to the negative side, the walk was continued by taking the absolute value of its position. The histogram of data as a probability density function is collected for $\alpha=1.5$. We have carried out the simulations for different initial positions $x=a$. The Fig. 1 depicts the results for $a=0$. The probability density function that we get from this simulation coincides with the analytic expression in Eq.~(\ref{eqR3}) as shown in Fig. 1a, and it also has power law behavior as $~x^{-(1+\alpha)}~$ for large $x$ (Eq.~(\ref{eqR5})) that matches with analytical data as seen in Fig. 1b. For comparison, the symmetric L\'evy distribution without the boundary is also shown in the figure.

In Fig. 2, we depict the results for three different nonzero values of initial positions $a$. In all these cases, the analytical and simulation results coincide, demonstrating that the setting of probability current to zero indeed provides the correct way to incorporate the reflecting boundary condition.

\section{Absorbing boundary}
In this section, we continue the discussion of the problem of L\'evy flight with restriction on the motion of the particle introduced by the presence of absorbing walls.  For normal diffusion, the absorbing boundary is incorporated by making the probability density zero at the boundary. Clearly, this is not sufficient for the L\'evy flights, and we have to explicitly demand that the probability density has to be zero outside the region of interest. 

We consider the semi-infinite domain $[0,\infty)$. The problem of L\'evy flights in a bounded domain with an absorbing boundary was considered in~\cite{MR} using the expansion in eigenfunctions. Clearly, this method cannot be used for a semi-infinite domain. We place the absorbing boundary walls at $x=0$. The absorbing boundary condition  would have to be incorporated by demanding that 
\begin{equation}\label{eqABC} 
 P(x,t)=0~~~ \forall~~~ x \leq 0.
\end{equation}
where $P(x,t)$ is the probability density. To apply this boundary condition on the symmetric L\'evy flights in $[0,\infty)$, we start with the Chapman-Kolmogorov equation. It is given by 
\begin{widetext}
\begin{equation}\label{eqAS1}
p(x-x_0;t-t_0)=\int_0^{\infty} p(x-x';t-t')p(x'-x_0;t'-t_0)dx'
\end{equation}
where the particle jumps from location $x_0$ to $x$ in time $t_0$ to $t$ and $x'$ is the intermediate location at time $t'$. Notice that the lower limit of integration is taken as zero. It is a consequence of the boundary condition as the intermediate location $x'$ cannot be less than 0.

Now we follow the steps in~\cite{benson} to arrive at the corresponding fractional differential equation. The time derivative of the probability density function $p$ is defined as
\begin{equation}\label{eqAS2}
\frac{\partial p(x-x_0;t)}{\partial t}= \lim_{\Delta t\to 0} \frac{1}{\Delta t}\left[p(x-x_0;t+\Delta t)-p(x-x_0;t)\right]
\end{equation}
Using the Chapman-Kolmogorov equation (\ref{eqAS1}) we can write
\begin{equation}
\begin{split}
p(x-x_0;t+\Delta t)&=\int_0^\infty p(x-x';\Delta t)p(x'-x_0;t)dx' \nonumber\\
&=\int_0^\infty p(x-x';\Delta t)P(x',t)dx' \nonumber
\end{split}
\end{equation}
where $p(x-x_0;t)=P(x,t)$. 
Substituting this value in the equation (\ref{eqAS2}), we get
\begin{eqnarray}\label{eqAS3}
\frac{\partial p(x-x_0;t)}{\partial t} &=& \lim_{\Delta t\to 0} \frac{1}{\Delta t}\left[\int_0^\infty p(x-x';\Delta t)P(x',t)dx'-P(x,t)\right]\nonumber\\
\frac{\partial P(x;t)}{\partial t} &=& \lim_{\Delta t\to 0} \frac{1}{\Delta t}\left[\int_0^\infty p(x-x';\Delta t)P(x',t)dx'-P(x,t)\right]
\end{eqnarray}
In order to arrive at the fractional differential equation governing this process, we note that the symmetric stable L\'evy transition density with zero mean, $p(x,t)$, has a Fourier transform given by  $\hat{p}(k,t)=e^{-|{ck}|^\alpha t}$, where $c$ is a scaling constant. We choose $c=[\cos(\frac{\pi}{2}(2-\alpha))]^{1/\alpha}$ and expand it as follows
\begin{equation}\label{eqAS4}
\hat{p}(k;\Delta t)=1+\frac{1}{2}\Delta t\left[(ik)^\alpha+(-ik)^\alpha\right]+o(\Delta t).
\end{equation}
\end{widetext}
The instantaneous transition density has the following limit:
\[\lim_{\Delta t\to 0} p(x;\Delta t)= \delta(x).\]
Therefore, we have
\[\lim_{\Delta t\to 0} \hat{p}(k;\Delta t)=1.\]
Taking the Fourier inverse transform of both sides of (\ref{eqAS4}), we arrive at
\begin{widetext}
\[p(x;\Delta t)=\delta(x)+\frac{\Delta t}{2}[_{-\infty}D_{x}^\alpha \delta(x)+_xD_{\infty}^\alpha \delta(x)]+o(\Delta t)\]
%Using the properties of the delta function, we get
%\[ D_{+}^\alpha \delta(x)= \frac{1}{\Gamma(2-\alpha)}\frac{\partial^2}{\partial x^2}\left[ \int_{-\infty}^x (x-y)^{1-\alpha} \delta(x-y)dy \right]\] and \[ D_{-}^\alpha \delta(x)=\frac{1}{\Gamma(2-\alpha)}\frac{\partial^2}{\partial x^2}\left[ \int_x^\infty (y-x)^{1-\alpha} \delta(x-y)dy \right].\] 
Substituting this above representation in the equation (\ref{eqAS3}), using the properties of the Dirac delta function and imposing the absorbing boundary condition $~P(x,t)=0$ for $x\leq 0$,   we get the following equation
\begin{eqnarray}\label{eqAS7}
\hspace*{-0.8in}\frac{\partial P(x,t)}{\partial t} &=& \lim_{\Delta t\to 0} \frac{1}{\Delta t} \left[P(x,t)+\frac{\Delta t}{2}\left[_{-\infty}D_{x}^\alpha P(x,t)+_xD_{\infty}^\alpha  P(x,t)\right]+o(\Delta t)-P(x,t)\right]\nonumber\\
&=& \frac{1}{2} \left[ _{-\infty}D_{x}^\alpha P(x,t)+_xD_{\infty}^\alpha P(x,t)\right] \nonumber\\
&=& \frac{1}{2\Gamma(2-\alpha)}\frac{\partial^2}{\partial x^2} \left[ \int_{0}^x (x-\xi)^{1-\alpha}P(\xi,t)d\xi +\int_x^\infty (\xi-x)^{1-\alpha}P(\xi,t)d\xi \right].
\end{eqnarray}
So we have to solve
\begin{equation}\label{eqAS8}
\frac{\partial P(x,t)}{\partial t} = \frac{1}{2\Gamma(2-\alpha)}\frac{\partial^2}{\partial x^2} \left[ \int_{0}^x (x-\xi)^{1-\alpha}P(\xi,t)d\xi + \int_x^\infty (\xi-x)^{1-\alpha}P(\xi,t)d\xi \right] 
\end{equation}
and demand that $P(0,t)=0$.
To solve this equation (see Appendix C) we use the Fourier transform with respect to `x' and use the scaling property (see, for example,~\cite{dubkov}). We take the initial condition at $t=0$, $~P(x,0)=\delta(x-a)$ where initially the particles are on $x=a$, $a\geq 0$ to solve equation (\ref{eqAS8}). % We also make use of the Miller-Ross function \cite{miller, mainardi} to find the inverse Laplace transform and also of modified incomplete gamma function. 
We finally obtain the following.

\begin{equation}\label{eqAS9}
 P(x,t) = \frac{1}{\pi}\int_0^\infty e^{-\tilde{A}k^\alpha t} \cos(k(x-a)) dk -\frac{C}{2\pi\Gamma(2-\alpha)}\int_0^\infty E_i(\tilde{A}k^\alpha t) e^{-\tilde{A}k^\alpha t} \cos(kx)dk
\end{equation}
 where $ E_i $  is the exponential integral function  and $\tilde{A}=\cos(\frac{\pi}{2}(2-\alpha))$.
\end{widetext}

We have not yet used the boundary condition at $x=0$, which can be used to determine $C$. Therefore, demanding that $P(0,t)=0$ at $~~x=0~~$ for all $t$, we find the value of $C$ as:
\[ C={2\Gamma{(2-\alpha)}}\frac{\int_0^\infty e^{-\tilde{A}k^\alpha t} \cos(ka)dk}{\int_0^\infty E_i(\tilde{A}k^\alpha t) e^{-\tilde{A}k^\alpha t}dk}\]

Following a similar method as discussed in Appendix B and using the result of series expansion of the exponential integral function \cite{stegun}, we derive the series solution for $P(x,t)$ as follows.
\begin{widetext}
\begin{align}\label{eqAS11}
P(x,t)&=\frac{1}{\pi} \sum_{m = 1}^{\infty} \frac{(-1)^{3m+1}}{(2m-1)!} t^{-\frac{(2m-1)}{\alpha}} \tilde{A}^{-\frac{(2m-1)}{\alpha}}  \Bigg[ \Bigg.   \Gamma\left(1+\frac{(2m-1)}{\alpha}\right) \sum_{r = 0}^{2(m-1)} (-1)^r  \binom{2(m-1)}{r} a^rx^{2(m-1)-r}\nonumber\\
&- \frac{C}{2\Gamma{(2-\alpha)}}  \biggl\{  \left(\gamma_E+\psi(\frac{2m-1}{\alpha}) \right)  \Gamma\left(1+\frac{2m-1}{\alpha}\right) + \sum_{n = 1}^{\infty} \frac{(2m-1)\Gamma(n+\frac{2m-1}{\alpha})}{\alpha n \Gamma(1+n)}\biggl\}  x^{2(m-1)}   \Bigg. \Bigg]
\end{align}
where $ \gamma_E $  is the Euler constant  and  $\psi(x)$ is the logarithmic derivative of the gamma function or the digamma function.
\end{widetext}

\begin{figure}[ht]
          \centering
          \includegraphics[width=1.0\columnwidth]{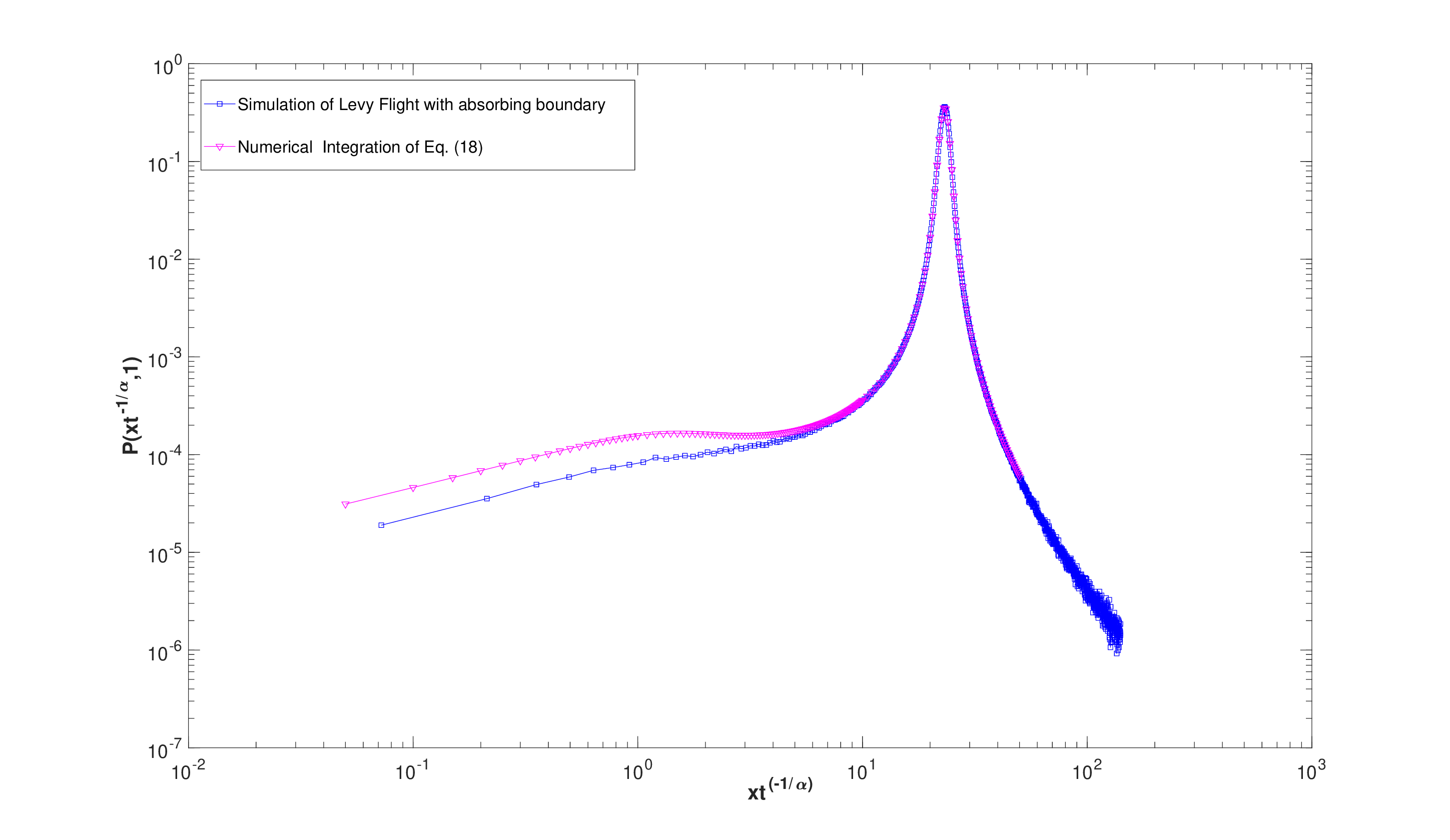}     
\caption{ Comparison of the solution obtained by numerical integration of equation (\ref{eqAS9}) and the simulation of the probability distribution function $P(x,t)$  of L\'evy flight with absorbing boundary for the value $~\alpha=1.5~$ and $~a=23.21~$ on  log-log scale. There is a good match between the two for larger values on the abscissa.}
\label{fig:absorbing}
      \end{figure}

Fig.~(\ref{fig:absorbing}) shows the graph of the analytical solution given by equation (\ref{eqAS9}) and the simulation of the probability distribution function $P(x,t)$ of the L\'evy flight with absorbing boundary for the value $~\alpha=1.5~$ and $~a=23.21~$ on the log-log scale. For simulation, we use GSL Library function (equation (10) with $c=[\cos(\frac{\pi}{2}(2-\alpha))]^{1/\alpha}$) for generating the random walk with jumps according to the L\'evy distribution, and whenever the walker jumps to the left of 0, we restart the walk from $x=a$. There is a small mismatch for lower values of $x$ possibly due to the assumption of the scaling relation while solving the equation. This point will be explored further in the future. 

\section{Conclusion}

In this work, we have considered L\'evy flights in a semi-infinite domain with reflecting and absorbing boundary conditions. To the best of our knowledge, this is the first exact treatment of a L\'evy flight with a boundary condition. 
Though L\'evy flights have been used to model many stochastic jump processes with the probability of jumps decaying as a power law, it has not been possible to use it in systems where the boundaries are nearby. A systematic approach to imposing boundary conditions was lacking. The problem was tricky, as the local boundary conditions employed for the usual random walks fail in the case of L\'evy flights owing to the long jumps involved.

We have demonstrated how reflecting and absorbing boundaries can be included in the equations governing the L\'evy motion and obtain expressions for the probability densities for a case of semi-infinite domain. For the problem with a reflecting boundary, we write a continuity equation and identify the expression for the probability current. We then equate this to zero at the reflecting boundary. To deal with the absorbing boundary, we start with the Chapman-Kolmogorov condition and then derive the governing equation by demanding that the probability density is zero outside the domain of interest. We have managed to solve the resulting equations analytically for symmetric L\'evy flights with the L\'evy index between 1 and 2 under the assumption that it satisfies a scaling relation. We have also compared our analytical results with Monte Carlo simulations of the L\'evy flights with reflecting and absorbing boundaries. The analytical and simulation results agree for all values in the case of reflecting boundary. In the case of absorbing boundary, there is a mismatch for smaller values of abscissa possibly due to the assumption of scaling while solving the fractional differential equation. This point needs to be examined in the future.

With the methods we have used, it should also be possible to extend our calculations to the other cases. The efforts are underway to generalize the results to the other range of L\'evy index and also to asymmetric L\'evy flights. Considering bounded domains~\cite{MR} would also be interesting from the point of view of applications to physical systems. 

The probability densities thus obtained would lead to the other quantities such as moments, first or mean passage times~\cite{SM}, which would have a direct physical relevance.

\begin{acknowledgments}
    This project has been financed by Council of Scientific and Industrial Research, India (03(1357)/16/EMR-II) and the Department of Science and Technology (DST), India under the Women Scientists Scheme-A (WOS-A) for Research in Basic/ Applied Sciences. We thank the anonymous referees for useful comments that improved the manuscript.
\end{acknowledgments}

\appendix

\begin{widetext}
\section{Derivation of solution of the Fractional Differential equation (\ref{eqR1}) with reflecting boundary }
We use the Fourier cosine transform defined as 
\begin{equation}\nonumber
\mathcal{F}_c\{f(x)\}=\sqrt{\frac{2}{\pi}}\displaystyle\int_0^\infty f(x) \cos(kx)dx.
\end{equation}
Taking the Fourier cosine transform of both sides of equation (\ref{eqR1}) with respect to `x', we obtain
\begin{equation*}
\frac{\partial}{\partial t} [ \mathcal{F}_c\{P(x,t)\}] =\mathcal{F}_c\left\{\frac{1}{\Gamma(2-\alpha)}\frac{\partial^2}{\partial x^2}  \left[\displaystyle\int_{-\infty}^x \frac{P(y,t)}{(x-y)^{\alpha-1}}dy + \displaystyle\int_x^\infty \frac{P(y,t)}{(y-x)^{\alpha-1}}dy \right] \right\} 
\end{equation*}
Since, \(\mathcal{F}_c\{f''(x)\} =-k^2 \mathcal{F}_c\{f(x)\}-\sqrt{\frac{2}{\pi}}f'(0)  \),
we have 
\begin{eqnarray}\frac{\partial}{\partial t} [ \mathcal{F}_c\{P(x,t)\}] &=&-k^2\mathcal{F}_c\left\{\frac{1}{\Gamma(2-\alpha)} \left[\displaystyle\int_{-\infty}^x \frac{P(y,t)}{(x-y)^{\alpha-1}}dy  + \displaystyle\int_x^\infty \frac{P(y,t)}{(y-x)^{\alpha-1}}dy  \right] \right \} \nonumber\\
 &&\;\;\;\;\;\;\;\;\;\;\;\;\;\;\;- \sqrt{\frac{2}{\pi}} \frac{\partial}{\partial x} \left\{ \frac{1}{\Gamma(2-\alpha)} \left[ \displaystyle\int_{-\infty}^x \frac{P(y,t)}{(x-y)^{\alpha-1}}dy + \displaystyle\int_x^\infty \frac{P(y,t)}{(y-x)^{\alpha-1}}dy  \right] \right\} \bigg\rvert_{x=0}. \nonumber
\end{eqnarray}
%\intertext{\centering[using the boundary condition given by equation $(\ref{RBC1})$ we get]}
Using the boundary condition given by equation (\ref{RBC1}), we get
\begin{equation}\label{eqR7}
\hspace*{-1.4in}\frac{\partial}{\partial t} [ \mathcal{F}_c\{P(x,t)\}] =-k^2\mathcal{F}_c\left\{\frac{1}{\Gamma(2-\alpha)} \left[ \displaystyle\int_{-\infty}^x \frac{P(y,t)}{(x-y)^{\alpha-1}}dy + \displaystyle\int_x^\infty \frac{P(y,t)}{(y-x)^{\alpha-1}}dy \right]\right \} 
\end{equation}
The convolution formula of Fourier cosine transform  \cite{nathnath} is given by 
\begin{equation}\label{eqR8}
\begin{split}
(f*g)(x) & =\frac{1}{\sqrt{2\pi}}\displaystyle\int_0^\infty f(y)[g(|x-y|)+g(x+y)]dy\\ \nonumber
&=\frac{1}{\sqrt{2\pi}}\displaystyle\int_0^x f(y)g(x-y) dy + \frac{1}{\sqrt{2\pi}}\displaystyle\int_x^\infty f(y)g(y-x) dy+\frac{1}{\sqrt{2\pi}}\displaystyle\int_0^\infty f(y)g(x+y)dy\\ 
\intertext{Considering f(y) as an even function, the formula reduces to}
&=\frac{1}{\sqrt{2\pi}}\displaystyle\int_{-\infty}^x f(y)g(x-y) dy+\frac{1}{\sqrt{2\pi}}\displaystyle\int_x^\infty f(y)g(y-x)dy
\end{split}
\end{equation}
Again defining $ g(y)=y^{1-\alpha}$, $y\geq 0~$ and applying above convolution formula, the equation (\ref{eqR7}) can be rewritten  as 
\begin{equation} \label{eqR13}
%\begin{split} 
\frac{\partial}{\partial t} [ \mathcal{F}_c\{P(x,t)\}] = -k^2\mathcal{F}_c\left\{\frac{\sqrt{2\pi}}{\Gamma(2-\alpha)} (P*g)(x) \right\} = -k^2\frac{\sqrt{2\pi}}{\Gamma(2-\alpha)} \mathcal{F}_c\left\{ P(x,t) \right\} \mathcal{F}_c\left\{g(x) \right\}
%\end{split}
\end{equation}
Now we use the following result from \cite{erdelyi,nathnath} $$\mathcal{F}_c\{g(x)\} =\sqrt{\frac{2}{\pi}}\displaystyle\int_0^\infty x^{1-\alpha}\cos(kx)dx  =\sqrt{\frac{2}{\pi}} \frac{\Gamma(2-\alpha)}{k^{2-\alpha}}\cos(\frac{\pi}{2}(2-\alpha))$$ 
in the equation (\ref{eqR13}) to obtain 
\begin{eqnarray} \label{eq15}
\frac{\partial}{\partial t} [ \mathcal{F}_c\{P(x,t)\}] &=& -k^2\frac{\sqrt{2\pi}}{\Gamma(2-\alpha)}\sqrt{\frac{2}{\pi}} \Gamma(2-\alpha)k^{\alpha-2}\cos(\frac{\pi}{2}(2-\alpha))\mathcal{F}_c\{P(x,t)\} \nonumber \\
&=& -Ak^\alpha\mathcal{F}_c\{P(x,t)\}\mbox{~~~~~where $A=2\cos(\frac{\pi}{2}(2-\alpha))$} 
\end{eqnarray}
 Integrating, we get,
\[ \mathcal{F}_c\{P(x,t)\} =\mathcal{F}_c\{P(x,0)\} e^{-Ak^\alpha t} \]

Now, we consider the initial condition $P(x,0)=\delta(x-a)$ (Dirac delta function). Therefore, $\mathcal{F}_c\{P(x,0)\} = \mathcal{F}_c\{\delta(x-a)\}=\sqrt{2/\pi}\cos(ka)$ and
$$\mathcal{F}_c\{P(x,t)\}=\sqrt{\frac{2}{\pi}} e^{-Ak^\alpha t}\cos(ka) $$
Taking Fourier cosine inverse transform, we restore the value of $P(x,t)$ as
\begin{equation}\label{eqR17}
\begin{split}
P(x,t)  &= \sqrt{\frac{2}{\pi}}\displaystyle\int_0^\infty \sqrt{\frac{2}{\pi}} e^{-Ak^\alpha t} \cos(ka)\cos(kx)dk \\
 &= \frac{2}{\pi}\displaystyle\int_0^\infty e^{-Ak^\alpha t} \cos(ka) \cos(kx)dk
\end{split}
\end{equation}

%****************************************************************************************************************************************
\section{Derivation of series solution from the integral (\ref{eqR2}) by using Mellin transformation}
We have
\begin{equation}
P(x,t) =  \frac{2}{\pi}\displaystyle\int_0^\infty e^{-A k^\alpha t} \cos(ka)\cos(kx) dk\\
\end{equation}
Now substitute $k=ak'$ and $a=t^{-\frac{1}{\alpha}}$, i.e. $a^\alpha t=1$. Then $P(x,t)$ can be represented in this form \[P(x,t) = t^{-\frac{1}{\alpha}}K(x/t^{\frac{1}{\alpha}})\] where
\begin{align}
K(x/t^{\frac{1}{\alpha}})&=  \frac{2}{\pi} \displaystyle\int_0^\infty e^{-A k^\alpha } \cos(ka/t^{\frac{1}{\alpha}})\cos(kx/t^{\frac{1}{\alpha}})dk \nonumber\\
&= \frac{1}{\pi} \displaystyle\int_0^\infty e^{-A k^\alpha } \left[\cos\left(k(x+a)/t^{\frac{1}{\alpha}}\right)+\cos\left(k(x-a)/t^{\frac{1}{\alpha}}\right)\right]dk \nonumber\\
&= K_1\left(k(x+a)/t^{\frac{1}{\alpha}}\right)+K_2\left(k(x-a)/t^{\frac{1}{\alpha}}\right) 
\end{align}
Mellin convolution formula  \cite{nathnath} is given by $H(r)=\displaystyle\int_0^\infty \frac{1}{\rho} f(\rho) g(r/\rho)d\rho$\\
Comparing the above equation with \[K_1(k(x+a)/t^{\frac{1}{\alpha}})=\frac{1}{\pi} \displaystyle\int_0^\infty e^{-A k^\alpha }\cos(k(x+a)/t^{\frac{1}{\alpha}})dk,\] we get $\rho=k$, $r=\frac{1}{|x+a|}$, $f(k)=e^{-A k^\alpha }$ and \[g(k)= \frac{1}{\pi}\frac{1}{k|x+a|} \cos(\frac{1}{kt^{\frac{1}{\alpha}}}).\]
Therefore 
\begin{eqnarray}
K_1((x+a)/t^{\frac{1}{\alpha}}) &=& \frac{1}{\pi} \displaystyle\int_0^\infty e^{-A k^\alpha } \cos(k(x+a)/t^{\frac{1}{\alpha}})dk \nonumber \\
&=& \displaystyle\int_0^\infty \frac{1}{k} f(k) g(\frac{1}{k|x+a|})dk \nonumber \\
&=& f(k)*g(k) \nonumber
\end{eqnarray}
Taking the Mellin Transform on both sides with respect to `x' we get 
\begin{equation}\label{eqR20}
M\{K_1((x+a)/t^{-\frac{1}{\alpha}}\}=M\{f(k)*g(k)\}=f^*(s)g^*(s)
\end{equation}
Now \(f^*(s) =M\{f(k)\} =M\{e^{-A k^\alpha } \} =\frac{1}{\alpha} A^{-\frac{s}{\alpha}} \Gamma\left(\frac{s}{\alpha}\right) \)
and
\begin{eqnarray}
g^*(s) &=& M\{g(k)\} \nonumber \\
&=& M\left\{ \frac{1}{\pi}\frac{1}{k|x+a|} \cos(\frac{1}{kt^{\frac{1}{\alpha}}})\right\} \nonumber \\
&=& \frac{1}{\pi}\frac{1}{|x+a|}(t^{-\frac{1}{\alpha}})^{(s-1)} \Gamma(1-s) \cos\frac{\pi(1-s)}{2} \nonumber
\end{eqnarray}
Therefore, equation (\ref{eqR20}) becomes
\begin{align}\label{eqR23}
M\{K_1((x+a)/t^{\frac{1}{\alpha}})\} &= \frac{1}{\pi}\frac{1}{|x+a|}\frac{1}{\alpha} t^{\frac{1-s}{\alpha}} A^{-\frac{s}{\alpha}}\Gamma\left(\frac{s}{\alpha}\right)\Gamma(1-s) \cos\frac{\pi(1-s)}{2}. \nonumber\\
\intertext{ Taking both side $M^{-1}$ with respect to $x$, we get }
K_1((x+a)/t^{\frac{1}{\alpha}}) &=  \frac{1}{2\pi i}\frac{1}{\pi\alpha} \displaystyle\int_{\gamma-i\infty}^{\gamma+i\infty} t^{\frac{1-s}{\alpha}}  A^{-\frac{s}{\alpha}}\Gamma\left(\frac{s}{\alpha}\right)\Gamma(1-s) \sin\left(\frac{\pi s}{2}\right) (x+a)^{s-1}ds. \nonumber\\
\intertext{Similarly, we have}
K_2((x-a)/t^{\frac{1}{\alpha}}) &=  \frac{1}{2\pi i}\frac{1}{\pi\alpha} \displaystyle\int_{\gamma-i\infty}^{\gamma+i\infty} t^{\frac{1-s}{\alpha}}  A^{-\frac{s}{\alpha}}\Gamma\left(\frac{s}{\alpha}\right)\Gamma(1-s) \sin\left(\frac{\pi s}{2}\right) (x-a)^{s-1}ds.
\end{align}
Therefore, we finally have,
\begin{equation}\label{eqR24}
P(x,t) =t^{-\frac{1}{\alpha}}K(x/t^{\frac{1}{\alpha}}) = \frac{1}{2\pi i}\frac{1}{\pi \alpha} \displaystyle\int_{\gamma-i\infty}^{\gamma+i\infty} t^{-\frac{s}{\alpha}}  A^{-\frac{s}{\alpha}}\Gamma\left(\frac{s}{\alpha}\right)\Gamma(1-s) \sin\left(\frac{\pi s}{2}\right) \left[(x+a)^{s-1}+(x-a)^{s-1}\right] ds 
\end{equation}
where $\gamma=\mathbb{R}(s)$ and $0<\gamma<1$

The fundamental result on the Mellin Barnes integral~\cite{paris} assures us for $1<\alpha<2$ that the above contour of the integration can be transformed to the loop $L_{+\infty}$ starting and ending at $\infty$ and encircling all poles of the integral, given by $s=1+n,~~ n=0,1,2,\cdots$. 
Residue of the integral on RHS of equation (\ref{eqR24}) at $s=1+n$
\begin{align}\label{eqR25}
\hspace*{-1.0in}\mbox{Residue}\bigg\rvert_{s=1+n} &=\lim_{s \rightarrow 1+n} (s-(1+n))  t^{-\frac{s}{\alpha}} A^{-\frac{s}{\alpha}}\Gamma\left(\frac{s}{\alpha}\right)\Gamma(1-s) \sin \left(\frac{\pi s}{2}\right) \left[(x+a)^{s-1}+(x-a)^{s-1}\right]. \nonumber\\
\intertext{Using the result $\Gamma(1+z)=z\Gamma(z)$, we get}
& = {\alpha}\frac{(-1)^{n+1}}{(n+1)!} t^{-\frac{1+n}{\alpha}} A^{-\frac{1+n}{\alpha}}\Gamma\left(1+\frac{1+n}{\alpha}\right) \sin\left(\frac{(n+1)\pi}{2}\right)\left[(x+a)^{n}+(x-a)^{n}\right]  
\end{align}
Therefore, the series solution is given by
\begin{equation}\label{eqR26}
P(x,t) = \frac{1}{\pi}  \sum_{n = 1}^{\infty} \frac{(-1)^{n+1}}{n!} t^{-\frac{n}{\alpha}} A^{-\frac{n}{\alpha}}\Gamma\left(1+\frac{n}{\alpha}\right) \sin\left(\frac{n\pi}{2}\right) \left[(x+a)^{n-1}+(x-a)^{n-1}\right] 
\end{equation}

\section{Derivation of solution of the Fractional Differential equation (\ref{eqAS8}) with absorbing boundary }
We start with equation~(\ref{eqAS8}) and take Fourier transform on both sides. By integration by parts, we obtain
\begin{align}\label{eqAS12}  
\hspace*{-1.4in}\frac{\partial}{\partial t} [P(k,t)]&=- \frac{1}{\sqrt{2\pi}}\frac{k^2}{2\Gamma(2-\alpha)}\left[ \displaystyle\int_{0}^\infty \left\{\displaystyle\int_{0}^x \frac{P(\xi,t)}{(x-\xi)^{\alpha-1}}d\xi  + \displaystyle\int_x^\infty \frac{P(\xi,t)}{(\xi-x)^{\alpha-1}}d\xi  \right \}e^{ikx} dx + \displaystyle\int_{-\infty}^0 \displaystyle\int_x^\infty \frac{P(\xi,t)}{(\xi-x)^{\alpha-1}}d\xi e^{ikx} dx\right]\nonumber\\
 &+ \frac{1}{\sqrt{2\pi}}\frac{1}{2\Gamma(2-\alpha)} \left[- \frac{\partial}{\partial x} \left\{\displaystyle\int_{x}^\infty \frac{P(\xi,t)}{(\xi-x)^{\alpha-1}}d\xi \right\} \bigg\rvert_{x=0}  - \frac{\partial}{\partial x} \left\{\displaystyle\int_{0}^x \frac{P(\xi,t)}{(x-\xi)^{\alpha-1}}d\xi + \displaystyle\int_x^\infty \frac{P(\xi,t)}{(\xi-x)^{\alpha-1}}d\xi \right\} \bigg\rvert_{x=0} \right]
\end{align}
where $P(k,t)$ denotes the Fourier transform of $P(x,t)$.
After simplification, the first term of the equation (\ref{eqAS12}) becomes \[-k^{\alpha}\cos(\frac{\pi}{2}(2-\alpha)) P(k,t).\]
Using the scaling property~\cite{dubkov}, $P(\xi,t)=t^{-1/\alpha}P(\xi t^{-1/\alpha},1)=t^{-1/\alpha}P(\xi t^{-1/\alpha})$, the second term of the equation (\ref{eqAS12}) reduces to
\begin{align}\label{eqAS15}
\hspace*{-1.0in} &- \frac{\partial}{\partial x} \left\{\displaystyle\int_{x}^\infty \frac{P(\xi,t)}{(\xi-x)^{\alpha-1}}d\xi \right\} \bigg\rvert_{x=0}  - \frac{\partial}{\partial x} \left\{\displaystyle\int_{0}^x \frac{P(\xi,t)}{(x-\xi)^{\alpha-1}}d\xi + \displaystyle\int_x^\infty \frac{P(\xi,t)}{(\xi-x)^{\alpha-1}}d\xi \right\} \bigg\rvert_{x=0} \nonumber  \\ 
&= -t^{-1}\left[ \frac{\partial}{\partial y} \left\{\displaystyle\int_{0}^y \frac{P(z)}{(y-z)^{\alpha-1}}dz + \displaystyle\int_y^\infty \frac{P(z)}{(z-y)^{\alpha-1}}dz \right\} \bigg\rvert_{y=0}+\frac{\partial}{\partial y} \left\{ \displaystyle\int_{y}^\infty \frac{P(z)}{(y-z)^{\alpha-1}}dz \right\} \bigg\rvert_{y=0}\right] \nonumber\\
&= t^{-1} C \nonumber
\end{align}
where $C$ is a constant.
So the equation (\ref{eqAS12}) becomes 
\begin{equation}
\frac{\partial}{\partial t} P(k,t) + \tilde{A}k^\alpha P(k,t) = \frac{1}{\sqrt{2\pi}}\frac{C}{2\Gamma(2-\alpha)} t^{-1},   
\end{equation}
where $\tilde{A}=\cos(\frac{\pi}{2}(2-\alpha))]$
Solving this 1st order ODE with respect to `t', we get $P(k,t)$ in the form 
\[ P(k,t) = -\frac{1}{\sqrt{2\pi}}\frac{C }{2\Gamma(2-\alpha)}E_i(\tilde{A}k^\alpha t)e^{-\tilde{A}k^\alpha t}+Ee^{-\tilde{A}k^\alpha t}\]
where the exponential integral $ E_i $ is a special function in the complex plane, and E is the integration constant, whose value has to be found using the initial condition of the particles.

%$$ P(k,t) = \frac{\mathscr{D}C }{2\Gamma(2-\alpha)}\left[\log{(Ak^\alpha t)}+\sum_{n=1}^\infty \frac{(At)^n}{n\Gamma{(1+n)}}k^{n\alpha} \right]e^{-Ak^\alpha t}+Ee^{-Ak^\alpha t}$$
%where E is integration constant and whose value have to find using initial condition of particles.
At $t=0,~~ E=P(k,0)=\mathcal{F}\{P(x,0)\}=\mathcal{F}\{\delta(x-a)\}=\frac{1}{\sqrt{2\pi}}e^{ika}$. Using this value of $E$ and applying the inverse Fourier transform with respect to `k' we get $P(x,t)$ as 

\begin{equation}
\hspace*{-1.0in} P(x,t) = \frac{1}{2\pi}\int_{-\infty}^\infty e^{-\tilde{A}k^\alpha t} e^{ika}e^{-ikx} dk -\frac{C}{4\pi\Gamma(2-\alpha)}\int_{-\infty}^\infty E_i(\tilde{A}k^\alpha t) e^{-\tilde{A}k^\alpha t}e^{-ikx} dk
\end{equation}
Using the result from Mainardi \cite{mainardi2001} , $\psi(-k,\theta)=\psi(k,-\theta)$ , we get $ ~\psi(-k,\theta)=~\tilde{A}k^\alpha t=~cos(-\theta)k^\alpha t=~cos(\frac{\pi}{2}(\alpha-2))k^\alpha t=~cos(\frac{\pi}{2}(2-\alpha))k^\alpha t=\psi(k,\theta) $.

\begin{equation}
\hspace*{-1.0in} P(x,t) = \frac{1}{\pi}\int_0^\infty e^{-\tilde{A}k^\alpha t} \cos(k(x-a)) dk -\frac{C}{2\pi\Gamma(2-\alpha)}\int_0^\infty E_i(\tilde{A}k^\alpha t) e^{-\tilde{A}k^\alpha t} \cos(kx)dk
\end{equation}
\end{widetext}

%********************************************************************************************************************************************


\begin{thebibliography}{99}
\bibitem {zasla} M. F. Shlesinger, G. M. Zaslavsky, and J. Frisch, eds., L\'evy flights and related topics in physics ,(Springer Verlag, Berlin), (1995).
\bibitem{klafter1986} M. F. Shlesinger and J. Klafter, in On growth and form:Fractal and non-fractal patterns in physics, edited by H. E. Stanley and N. Ostrowsky (Springer Verlag, Berlin), 279 (1986).
\bibitem {metkla} R. Metzler and J. Klafter, The random walk's guide to anomalous diffusion: a fractional dynamics approach, Physics Reports, 339 , 1-77, (2000).
\bibitem {klafter96} J. Klafter, M. F. Shlesinger and G. Zumofen, Beyond Brownian Motion. Physics Today, 49, 33-39, (1996).
\bibitem {mandelbrot} B. Mandelbrot, The Fractal Geometry of Nature (Freeman, New York), (1977).
\bibitem{hufnagel2006} D. Brockmann, L. Hufnagel and T. Geisel, The scaling laws of human travel. Nature, 439(7075), 462–465, (2006).
\bibitem{hongchong} I. Rhee, M. Shin, S. Hong, K. Lee, S.J. Kim, S. Chong , On the Levy-Walk Nature of Human Mobility,IEEE/ACM Transactions on Networking, 19, 3, (2011).
\bibitem{klafter2015} V. Zaburdaev, S. Denisov, and J. Klafter, Levy Walk, Rev. Mod. Phys. 87, 483 (2015).
\bibitem{Catalan} F. Bartumeus, M. G. E. Da Luz, G. M. Viswanathan and J. Catalan, Animal search strategies: A quantitative random-walk analysis. Ecology 86, 3078–3087 (2005).
\bibitem{metzler} R. Metzler, A. V. Chechkin, and J. Klafter, “Levy statistics and anomalous transport: Levy flights and subdiffusion, ” in Computational Complexity: Theory, Techniques, and Applications, edited by A. R. Meyers (Springer New York, New York, NY, 2012) pp. 1724–1745.
\bibitem{corral} Corral, A. Universal earthquake-occurrence jumps, correlations with time, and
anomalous diffusion. Phys. Rev. Lett. 97, 178501 (2006).
\bibitem{costa} O. Sotolongo-Costa, J.C. Antoranz, A. Posadas, F. Vidal and A. Vfizquez, Levy Flights and Earthquakes , Geographycal Research Letters, 27, 13, 1965-1968,( 2000).
\bibitem{HH} H. Hakli and H. Uguz, A novel particle swarm optimization algorithm with Levy flight. Applied Soft Computing, 23, 333–345, (2014).
\bibitem{wiersma} P. Barthelemy, J. Bertolotti1 and D. S. Wiersma, A Levy flight for light, Nature , 453, 495-498, (2008).
\bibitem{geisel} T. Geisel, J. Nierwetberg and  A. Zacherel, Accelerated diffusion in Josephson junctions and related chaotic systems. Phys. Rev. Lett. 54, 616–619 (1985). 
\bibitem{sokolov} I. M. Sokolov, J. Mai, and A. Blumen, Phys. Rev. Lett. 79, 857 (1997).
\bibitem{ott} A. Ott, J. P. Bouchaud, D. Langevin, and W. Urbach, Phys. Rev. Lett. 65, 17 (1990).
\bibitem{hufnagel2004} L. Hufnagel, D. Brockmann, and T. Geisel, Proc. Natl. Acad. Sci. U.S.A. 101, 15124 (2004).
\bibitem{garbaczewski2018} P. Garbaczewski, Fractional Laplacians and Lévy Flights in Bounded Domains, Acta Physica Polonica B, 49, 5, 921, (2018).
\bibitem{garbaczewski2019} P. Garbaczewski and V. Stephanovich, Fractional Laplacians in bounded domains: Killed, reflected, censored and taboo Lévy flights, Phys. Rev. E, 99, 042126, (2019).
\bibitem{nowak2017} B. Dybiec, E. Gudowska-Nowak, E. Barkai and A. A. Dubkov, Lévy flights versus Lévy walks in bounded domains, Phys. Rev. E 95, 052102, (2017).
\bibitem{deng} W. Deng, B Li, W. Tian, and P. Zhang , Boundary Problems for the Fractional and Tempered Fractional Operators, Multiscale Model. Simul., 16(1), 125–149, (2018).
\bibitem{Araujo} H. A. Araújo and G. Pagnini, Pre-asymptotic analysis of Lévy flights,  Chaos 34, 073126 (2024).
\bibitem{dybiec} B. Dybiec, E. Gudowska-Nowak and P. Hanggi, Levy Brownian motion on finite intervals: Mean first passage time analysis, Phys. Rev. E 73, 046104 (2006).
\bibitem{SM} I. M. Sokolov and R. Metzler, Non-uniqueness of the first passage time density of
Lévy random processes,  J. Phys. A: Math. Gen. 37 L609 (2004).
\bibitem{chandrasekhar} S. Chandrasekhar, Stochastic problems in physics and astronomy, Reviews of Modern Physics, Vol. 15, 310 (1943). 
\bibitem{podlubny} I. Podlubny, Fractional differential equations, Mathematics in Science and Engineering , Vol 198, ISBN  0 -12 558840 -2, (1999).
\bibitem{schneider1} W. R. Schneider, Generalised one-sided stable distribution, Proceedings of the second BiBoS-Symposium. Albeverio, S., Blanchard, Ph., Streit, L.,(eds.). Lecture notes in Mathematics. Berlin: Springer (1986).
\bibitem{schneider2} W. R. Schneider, Stable distributions: Fox function representation and generalization, in Stochastic Processes in Classical and Quantum Systems, Lecture Notes in Physics, Vol. 262, edited by S. Albeverio, G. Casati and D. Merlini (Springer, Berlin, 1986).
\bibitem{mainardi2001}  F. Mainardi, Y. Luchko and G. Pagnini, The fundamental solution of the space-time fractional diffusion equation, Fractional Calculus and Applied Analysis, Vol 4,No2, 153 (2001).
\bibitem{mainardi2005} F. Mainardi, G. Pagnini, R. K. Saxena, Fox H functions in fractional diffusion, Journal of Computational and Applied Mathematics, 178 , 321 – 331, (2005).
\bibitem{MPG} F. Mainardi, G. Pagnini, R. Gorenflo, Mellin transform and
subordination laws in fractional diffusion processes, Fract. Cal.
Appl. Anal. 6,  441–459, (2003).
\bibitem{BK1} K. Bogdan and M. Kunze,  Stable processes with reflection, arXiv preprint arXiv:2410.03516 (2024).
\bibitem{BK2} K. Bogdan and M. Kunze, The fractional Laplacian with reflections, Potential Analysis, 61(2), 317-345 (2024).
\bibitem{oldhum} K. B. Oldham and J. Spanier, The Fractional Calculus: Theory and Applications of Differentiation and Integration to Arbitrary Order, Mathematics in Science and Engineering , Vol 111, ISBN 0-12-525550-0, (1974).
\bibitem{mainardi2007} F. Mainardi, G. Pagnini and R. Gorenflo, Mellin transform and subodination laws in fractional diffusion processes, arXiv preprint math/0702133. Feb 6, (2007).
\bibitem{mainardi2010} F. Mainardi, Fractional Calculus and Waves in Linear Viscoelasticity: An Introduction to Mathematical Models, Imperial College Press, (2010).
\bibitem{braaksma} B. L. J. Braaksma, Asymptotic expansions and analytic continuations for a class of Barnes integrals, Compositio Mathematics 15, 239 (1964).
\bibitem{paris} R. B. Paris and D. Kaminski, Asymptotics and Mellin-Barnes integrals, Encyclopedia of Mathematics and its Applications, 85, Cambridge University Press, (2001). 
\bibitem{bjwest} E. W. Montroll and B. J. West , On an enriched collection of stochastic processes, In: Fluctuation Phenomena (Montroll, E.W., and Lebowitz,
J.L., Eds.). North–Holland, Amsterdam, pp. 61–175 (1979).
\bibitem{benson} D. A. Benson, S. W. Wheatcarft and M. Meerschaert, The fractional-order governing equation of L\'evy motion, Water Resources Research, 36, 6, 1413 (2000).
\bibitem{nathnath} L. Debnath and D. Bhatta, Integral transforms and their applications, Chapman and Hall/CRC, (2016).
\bibitem{erdelyi} A. Erdelyi, W. Magnus , F. Oberhettinger and F. G. Tricomi, Tables of Integral Transforms, 2, McGraw-Hill, New York, NY, USA (1954). 
\bibitem{MR} D.A. Monroy and E.P. Raposo, Solution of the space-fractional diffusion equation on bounded domains of superdiffusive phenomena, Phys. Rev. E,110, 054119 (2024).
\bibitem{dubkov} A. A. Dubkov, B. Spagnolo and  V. V. Uchaikin, Levy flight superdiffusion: an introduction. International Journal of Bifurcation and Chaos, 18(09), 2649-2672(2008).
\bibitem{stegun} M. Abramowitz, I. A. Stegun and R. H. Romer, Handbook of Mathematical Functions with Formulas, Graphs, and Mathematical Tables, New York, NY, USA: Amer. Assoc. Phys. Teachers, (1988).

%\bibitem{kadanoff} Leo P. Kadanoff, Scaling and Multiscaling: Fractals and Multifractals,  Chinese Journal of Physics, 29,6, 613 (1991). 
%\bibitem{nakao} H. Nakao, Multi-scaling properties of truncated Lévy flights, Physics Letters A, 266(4-6), 282–289, (2000).
%\bibitem{miller} K. S. Miller and  B. Ross, An introduction to the fractional calculus and fractional differential equations, Wiley, (1993).

\end{thebibliography}
\end{document}